# DEVELOPMENT OF A MACHINE PROTECTION SYSTEM FOR THE SUPERCONDUCTING BEAM TEST FACILITY AT FERMILAB*

A. Warner[#], L. Carmichael, M. Church, R. Neswold, FNAL, Batavia, IL 60510, U.S.A

*Abstract*

Fermilab's Superconducting RF Beam Test Facility currently under construction will produce electron beams capable of damaging the acceleration structures and the beam line vacuum chambers in the event of an aberrant accelerator pulse. The accelerator is being designed with the capability to operate with up to 3000 bunches per macro-pulse, 5Hz repetition rate and 1.5 GeV beam energy. It will be able to sustain an average beam power of 72 KW at the bunch charge of 3.2 nC. Operation at full intensity will deposit enough energy in niobium material to approach the melting point of 2500 °C. In the early phase with only 3 cryomodules installed the facility will be capable of generating electron beam energies of 810 MeV and an average beam power that approaches 40 KW. In either case a robust Machine Protection System (MPS) is required to mitigate effects due to such large damage potentials. This paper will describe the MPS system being developed, the system requirements and the controls issues under consideration.

## INTRODUCTION

The beam at Fermilab's New Superconducting RF Beam Test Facility [1], when operational, will need systems to protect critical components from beam induced damages such as beam pipe collision and excessive beam losses. The MPS must therefore identify hazardous conditions and then take the appropriate action before damage is caused. Since the loss of a full bunch train can result in significant damage, the MPS must be able to interrupt the beam within a macro-pulse and keep the number of bunches below the damage potential once the protection system reacts; the goal is to keep the number of bunches on the order of 3-6 bunches. With the high possible bunch frequency of 3 MHz this necessitates a reaction time in the range of 1-2 μs with cable delay included for the 134 metre long machine. The MPS will use the status of critical sub-systems and losses measured by a fast Beam Loss Monitor (BLM) system, using scintillators and photomultiplier tubes (PMT) to identify potential faults. Once a fault is observed, the MPS can then stop or reduce beam intensity by removing the permit from different beam actuators, including the laser pulse controller.

## MACHINE LAYOUT

The machine layout is shown in Fig. 1. The accelerator consist of an electron gun, a 40 MeV injector, a beam acceleration section consisting of 3 TTF-type or ILC-type cryomodules initially, and multiple downstream beam lines for testing diagnostics and performing beam experiments. The facility will accommodate up to 6 cryomodules in the final stage. From a machine protection point of view, the dump locations shown describe the final destination of the beam that traverses a path along the beam-line. These paths are termed operation modes and are validated by the MPS before the beam permit is released. The MPS validates these paths by monitoring all critical devices and diagnostics along the path and ensuring that they are all in good status and ready to receive beam at the requested intensity.

The machine will be capable of operating over a wide range of beam parameters as long as the total beam power remains below the limit of the beam dump capability and satisfies radiation shielding requirements. For machine protection purposes several beam modes have been defined; the beam mode set limits on the number of bunches and therefore the intensity. Initially following two modes will be active in the system:

- Low intensity mode – which allows the minimal beam intensity needed for OTR/YAG diagnostics. This is below the threshold potential for beam induced damage. In this mode there is no fast reaction to beam loss within a bunch train.
- High intensity mode – which does not impose a limit on the number of bunches, but enables fast intra-train protection by the MPS.

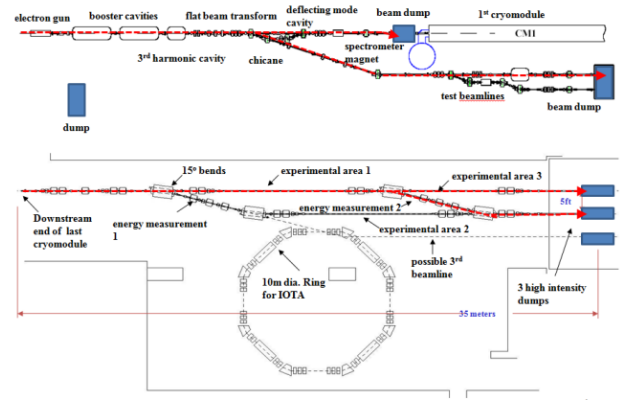

Figure 1: Machine layout

## MPS OVER-VIEW

The simplified overview block diagram of the proposed MPS is shown in Figure 2. The MPS has connections to several external devices and sub-systems. The top layer comprises signal providers such as fast beam loss monitors, RF signals, quench protection, toroid transmission, vacuum, magnet power supplies and more. All devices in this category send status information to the MPS logic layer (permit system). Only simple digital signals (e.g. on-off, OK-not OK) are transmitted. All devices or subsystems that are determined to be pertinent to protecting the machine or necessary for machine configuration are included here. The state of the machine is determined from this comprehensive overview of the inputs and allowable operational modes are determined based on this information by the middle logic layer. The main goals for the MPS system as a whole are:

- Provide precise protection of all critical components by first determining the fault severity (high, intermediate, etc) and then taking the appropriate action to avoid damage.
- Allow for high availability by ensuring that the maximum requested beam intensity is allowed for the detected fault severity.
- Monitor MPS components and perform periodic self-checks in order to ensure robustness and a high level of reliability.
- Provide well-integrated, user-friendly tools for fault visualization, control and post-mortem analysis.

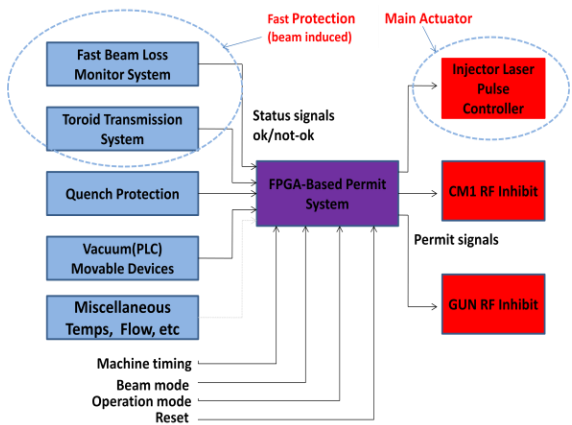

Figure 2: MPS Overview

The permit system part of the MPS is capable of handling events on all time scales relevant to the machine. This layer is FPGA based and is thus fully programmable and handles complex logic task. The logic here will be designed to ensure safe operating conditions by monitoring the status of critical devices and by imposing limits on the beam power. It prohibits beam production or reduces the beam intensity by disabling the gun RF and the injector laser unless the requirements for the specific predefined modes are fulfilled when that mode is requested. The final layer of the system shows the main actuators. This comprises of all points where the MPS logic may act on the operation of the machine and prevent beam from being produced or transported; the main actuator being the injector laser. When any of the non-masked inputs signal an alarm status the MPS permit system (logic layer) can do one of several things based on the severity of the fault: i.e. switch off the injector laser to suppress the production of new bunches, reduce the intensity by dialing back the number of bunches, or inhibit the rf power from the first cryomodule (CM1) as a precaution against transport of dark current from the RF gun.

## BEAM LOSS MONITOR SYSTEM

Several types of beam loss monitors (BLMs) will be used for the detection of electromagnetic showers. The fast protection system is being designed to interrupt the beam within a macro-pulse and will rely heavily on the ability to detect and react to losses within a few nanoseconds; for this reason the primary loss monitors for fast protection are made of plastic scintillator with photomultipliers attached and have already been designed, built and tested. Figure 3: shows some measurement results.

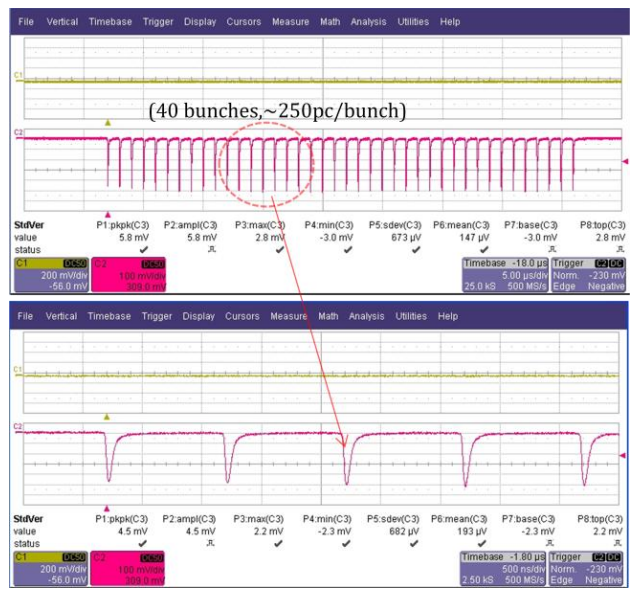

Figure 3: FBLM showing pulse beam losses

The BLMs will serve the dual purpose as an accelerator diagnostic and as the primary detectors for fast machine protection. The monitors must therefore deliver a measurement of beam and dark current losses to the control system as well as generate a fast alarm signal when the beam losses exceed user-defined thresholds. The time resolution of the loss measurement must provide the ability to distinguish single bunches within each macro pulse. This requires a sampling frequency of at least 3 MHz with a repetition rate of 5 Hz. The BLMs must be integrated into a robust beam loss monitoring system capable of generating an alarm condition that is derived by comparing the outputs of the PMT signals with various programmable thresholds. This alarm output is a critical component for machine protection. The desire is to provide a machine protection trip well before the beam can damage accelerator components. If one of the programmed thresholds is exceeded or if an error condition such as a high voltage failure or failed monitor is detected the system should report this to the MPS logic which in turn reduces the intensity or inhibits the beam.
The main requirements for the BLM system are:

- Provide both machine protection and diagnostic functions.
- Instantaneous read-back of beam loss
- Digital output for integrating and logarithmic signal (16 bit)
- Built in self test and onboard signal injection for testing of monitors between pulses.
- FPGA controlled
- Local data buffer
- VME interface to ACNET control system
- Continuous and pulsed monitoring
- Wide dynamic range

*Cryogenic Loss Monitors*

Although loss monitors are typically one of the main diagnostics for protecting the accelerator from beam induces damage. Most accelerator facilities do not cover the cold sections of the machine with loss monitors. To address these issues a Cryogenic Loss Monitor (CLM) ionization chamber capable of operation in the cold sections of a cryomodule has been developed and will be installed and tested [2]. The monitor electronics have been optimized to be sensitive to DC losses and the signals from these devices will be used to study and quantify dark current losses in particular, see figure 4. In order to increase the resolution bandwidth and the response time of the devices a new scheme which uses a Field Programmable Gate Array (FPGA) based Time-to-Digital converter (TDC) method is implemented [3] instead of a standard pulse counting method. This potentially renders these monitors as useful devices for both dark current monitoring and machine protection. These monitors under consideration are custom built detectors. They are an all metal designed which makes them intrinsically radiation hard and suitable for operation at 5 Kelvin to 350 Kelvin.

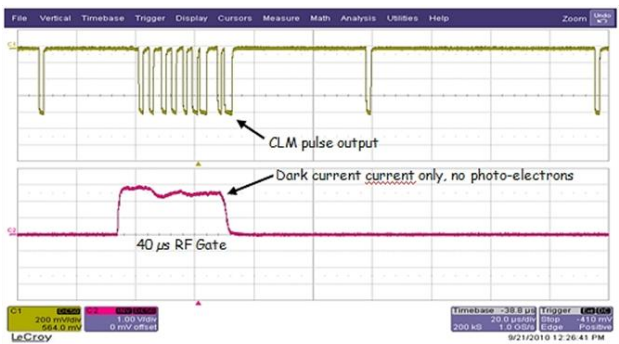

Figure 4: reaction of CLM to dark current losses in a test beam.

## LASER PULSE CONTROLLER

One of the main actuators for the MPS is the injector laser pulse control. This will be the device that controls the number and the spacing of bunches in a macro-pulse by picking single laser pulses out of a train. This is achieved by manipulating the Pockels cell (voltage-controlled wave plates). This system is also the main actuator for beam inhibits issued by the MPS. It is envisioned that this would be a VME board with a fully programmable FPGA. It would have inputs for the requested beam modes defined by the logic layer of the MPS, the MPS permit signal, the 3 MHz machine timing, and for a macro-pulse trigger. Based on the laser/gun design, it would have control outputs for the Pockels cell driver, a mechanical shutter and a first bunch timing signal. From the protection system point of view the pulse controller is used to:

- Block the Pockels cell based pulse kickers as long as the MPS input is in an alarm state.
- Enforce the limit on the number of bunches as given by the currently selected beam mode.
- Close the laser shutter on request of the MPS. This may happen when there is no valid operational mode or when some combination of loss monitors exceed thresholds which trigger a dump condition.

## CONTROLS INTEGRATION

The MPS will need server support for the various hardware systems to View, Configure and Diagnose the system. Already there are currently several servers under development for the beam loss monitor system and the laser pulse controller. These servers were implemented using the PowerPC 5500 series boards running VxWorks 6.4 and implementing the ACNET protocol. Some of the main requirements for these servers include:

- Time-stamping at a sub-microsecond resolution in order to allow for data correlation.
- Circular buffers that are logged using ACNET data loggers and thus provided a repository used for post-mortem analysis.

The integration of the control system with all MPS components, from the various front-ends to the high level applications, is critical to leveraging the full functionality of that control system.

## CONCLUSION

Significant effort is underway towards developing a reliable MPS for this new facility. System integration into the complex and commission challenges lay ahead.